# Microstructure and Conductance-Slope of InAs/GaSb Tunnel Diodes

Ryan M. Iutzi[a], Eugene A. Fitzgerald

*Department of Materials Science and Engineering, Massachusetts Institute of Technology, Cambridge, Massachusetts, 02139, USA*

InAs/GaSb and similar materials systems have generated great interest as a heterojunction for tunnel field effect transistors (TFETs) due to favorable band alignment. However, little is currently understood about how such TFETs are affected by materials defects and nonidealities. We present measurements of the conductance slope for various InAs/GaSb heterojunctions via two-terminal electrical measurements, which removes three-terminal parasitics and enables direct study on the effect of microstructure on tunnelling. Using this, we can predict how subthreshold swings in TFETs can depend on microstructure. We also demonstrate growth and electrical characterization for structures grown by metalorganic chemical vapor deposition (MOCVD) - a generally more scalable process compared to molecular beam epitaxy (MBE). We determine that misfit dislocations and point defects near the interface can lead to energy states in the band-gap and local band bending that result in trap-assisted leakage routes and nonuniform band alignment across the junction area that lower the steepness of the conductance slope. Despite the small lattice mismatch, misfit dislocations still form in InAs on GaSb due to relaxation as a result of large strain from intermixed compositions. This can be circumvented by growing GaSb on InAs, straining the GaSb underlayer, or lowering the InAs growth temperature in the region of the interface. The conductance slope can also be improved by annealing the samples at higher temperatures, which we believe acts to annihilate point defects and average out major fluctuations in band alignment across the interface. Using a combination of these techniques, we can greatly improve the steepness of the conductance slope which could result in steeper subthreshold swings in TFETs in the future.

---

[a] electronic mail: iutzi@mit.edu

## I. Introduction

Recently, tunnel field effect transistors (TFETs) have been the subject of much investigation due to their predicted and demonstrated ability to obtain subthreshold swings steeper than the room-temperature thermal limit of 60 mV/decade, as a result of interband tunneling[1,2]. Most early published results utilized heavily doped homojunctions and barrier-width modulation. However, recent simulation results have predicted that the use of heterojunctions allow for a much higher drive current, up to the order of mA/µm, due to the much smaller effective band gap and hence, higher tunnel probability[3-8]. In addition, devices from heterojunctions are expected to have a turn-on resulting from the band-edges overlapping as opposed to barrier width modulation. Conductance increases sharply once there is band overlap between the conduction band of one layer with the valence band of the adjacent layer, and each band edge provides a sharp cutoff in density of states. This removes the effect of thermal tails. However, while this concept is promising in theory, it remains to be seen if sharp switching can occur in a real heterojuction system with realistic imperfections.

Various researchers have predicted sharp switching from the band-edges in the InAs/GaSb-based materials system, due to its type-III band alignment, allowing for interband tunneling without heavy doping and with a high tunnel probability[4-8]. These computational results showed room-temperature subthreshold slopes as low as 7 mV/decade[8] and in some designs, drive currents as high as 1.9 mA/µm at 0.4 V [6]. Experimental results to date on this materials system[9-14], however, have not yet matched this performance, with the steepest room-temperature subthreshold slope at 125 mV/decade, and highest obtained drive current of 180 µA/µm at 0.5 V[12] at this time of writing. The reasons for this discrepancy have been hypothesized to result from a variety of factors, such as poor gate oxide interface[9-13], contact and other series resistance[10,11], TFET geometry[10,11,13], as well as materials defects related to the tunnel junction[9,10,12,14]. Furthermore, even in a two terminal configuration, where the effects of gate oxide and electrostatic coupling are removed, and with geometry expected to present less of an effect, work to date on two-terminal devices for this materials system[15-24] has also not resulted in electrical characteristics that correspond to switching steeper than the thermal limit[25]. This is a strong indication that materials effects could be the source of poor performance. Hence, understanding a link between materials properties and electrical results is imperative. This work addresses this need.

Additionally, experimental work on tunnel devices utilizing the InAs/GaSb system have been mainly based on layer structures formed with molecular beam epitaxy (MBE) growth[15-24], with very limited work with metalorganic chemical vapor deposition (MOCVD)[26], which has the promise to be a much more scalable technique, and hence useful for production of post-CMOS integrated circuits. There has been previous work, however, in this materials system for type-II superlattice long-wavelength detectors[27-31], as well as some early structure-oriented InAs/GaSb heterojunction work[32]. These studies have elucidated to the presence of various types of materials issues, including strain buildup and dislocation formation[29], as well as intermixing due to exchange and/or gallium carryover[28-30,32-34]. While the gallium carryover effect seems to be specific to MOCVD, dislocations[14,35,36] and exchange[37,38] are also seen in MBE. There has been no correlation to date on how defects on MOCVD-grown structures could affect tunneling properties. There has also been little work on this correlation for MBE-growth, although some recent results by Zhu et al. have identified a link between dislocations due to strain buildup and a resulting high off-state current[14,35]. These are the first results to explicitly show a major effect of growth properties on device performance and indicates the importance of linking materials structure to electrical properties.

Herein, we present our work on MOCVD-grown InAs/GaSb structures and the resultant structural properties linked to growth conditions, such as temperature, growth order, and strain. Finally, we have fabricated tunnel diodes from these epitaxial structures in order to obtain a device that can probe the tunneling characteristics without convoluting the result with the various transistor effects mentioned prior. We correlate the electrical properties with the materials properties to gain insights into how these devices are affected by materials defects, both those specific to MOCVD and also technique-independent results.

## II. Theory and Experiment

### A. Two-Terminal vs. Three-Terminal Device Characterization

A TFET is a three-terminal device and its subthreshold swing is a figure of merit that is specific to three-terminals. However, studying the effect of microstructure on the subthreshold swing of a TFET is difficult because it also depends heavily on other device parasitic issues that cannot easily be controlled or deconvoluted from the final electrical characteristics. These parasitics include interface states at the oxide-semiconductor interface, series resistance, and a loss of symmetry in the device geometry in going from two to three terminals, which results in nonuniform electrostatics and hence, parasitic leakage pathways. While it is important to optimize these aspects of the device, a study that focuses directly on the microstructure at the tunnel interface would benefit greatly from removing the guesswork of correcting for these variables.

A diode measurement provides an advantage in that it does not have as many parasitic effects. Namely, there is no gate-oxide that needs to be controlled, and there are no parasitic leakage pathways resulting from a third-terminal. Further, a diode demonstrates a two-terminal equivalent to a subthreshold swing through the dependence of its absolute conductance on the applied voltage. This was demonstrated recently by Agarwal[25], in which it was shown that a plot of the absolute conductance as a function of voltage gives a pseudo-transistor response.

The tunnel current of a tunnel diode ($J_t$) can be described by:

$$J_t = C_1 \int^{\Delta E} [F_C(E) - F_V(E)] T_t N_C(E) N_V(E) \, dE \quad (1)$$

where $F_c(E)$ and $F_v(E)$ are the Fermi-Dirac distributions on either side of the tunnel junction, $T_t$ is the tunnel probability, $N_c(E)$ and $N_v(E)$ are the density of states in the conduction band and valence bands respectively. $C_1$ is a constant, and the integral is taken over the band overlap $\Delta E$. In practice, as will be described in the next section, the $N_c(E)$ and $N_v(E)$ terms do not suddenly decrease to zero at the band edges, but are blurred due to energy states below the band edges and will also have a spatial dependence, meaning that $J_t$ will not be uniform everywhere across the junction. The applied potential ($V_a$) shifts the band overlap by shifting $N_c(E)$ and $N_v(E)$ away from each other, but also changes the difference in Fermi levels across the junction, $F_C(E) - F_V(E)$. However, in a three-terminal device, a gate would determine the band overlap without setting up a difference in Fermi-level across the junction. This results from the fact that an ideal gate causes no current to flow across the oxide. While the gate does induce a charge in the channel, the electric potential resulting from it is offset by the chemical potential difference, and the Fermi-level, which is the sum of these two potentials, remains constant. Hence, the entire Fermi-level difference ($qV_g$) is dropped across the oxide. The difference in Fermi-level across the junction is then controlled by the source-drain bias, and so the $F_C(E) - F_V(E)$ term is independent of gate bias and not part of the transfer characteristics and subthreshold slope of the transistor.

As a result, in order to approximate the equivalent scenario in a diode, one needs to remove the $[F_C(E) - F_V(E)]$ factor from the integrand. To first order, this difference in Fermi-distribution across the junction is simply the portion of the applied voltage that is dropped across the junction (this assumption will be validated at the end of the section). Therefore, this effect of Fermi-level difference can be removed by examining the absolute conductance of the junction:

$$G_t = \frac{J_t}{V_a} \approx C_1 \int^{\Delta E} TN_C(E)N_V(E)\,dE \quad (2)$$

And therefore, defining a conductance slope in a diode gives a comparable metric to the expected subthreshold slope if the diode were converted to a three-terminal configuration without the accompanying three-terminal parasitics:

$$Conductance\ Slope = \frac{dV_a}{dlog(G_t)} \quad (3)$$

Hence, the conductance slope allows one to see the subthreshold swing that could be obtained without three-terminal parasitics. Also, the series resistance can be easily determined and corrected by examining the slope of the I-V curve at high bias when the diode is essentially a series resistance-limited p-n junction.

There are a few important caveats that must be considered during analysis of the conductance slope.

1. To first order, the $[F_C(E) - F_V(E)]$ term in (1) only scales the integral by $V_a$ and hence can be divided out. However, to higher order, $[F_C(E) - F_V(E)]$ has an energy-dependence because of the thermal-tails of the Fermi-functions, and so it also weighs the integrand by the value of $[F_C(E) - F_V(E)]$. Below an applied bias of 4kT (100 mV), the weighting barely changes with bias, since the energy dependence of the $[F_C(E) - F_V(E)]$ term is dictated by the thermal tails of $F_C(E)$ and $F_V(E)$, and so the weighting of the $[F_C(E) - F_V(E)]$ term does not affect the conductance slope. Beyond an applied bias of 4kT, the energy dependence of the $[F_C(E) - F_V(E)]$ term is also affected by the applied bias, and the weighting changes. However, as Agarwal has shown[25], in the range of voltages from which we extract a conductance slope in this work (0 to 8kT/200 mV), the $[F_C(E) - F_V(E)]$ energy dependence at any given energy does not change by more than a factor of 2, and its total integral over energy is always equal to $V_a$. When integrated with the $N_C(E)N_V(E)$ term, which changes by many orders of magnitude, this factor of 2 becomes insignificant, and is not expected to change the calculated value of the conductance slope.

2. Tunnel diodes with negative differential resistance (NDR) tend to have instability-driven oscillations in and near the NDR region. Therefore, the absolute value of the conductance slope is meaningless in this region. However, as will be seen in the results, the turnoff begins before the NDR region, and so the high-current portion of the conductance slope can be seen without interference from oscillations. Furthermore, the average can be taken across the instability region to obtain an average conductance slope for that region, as has been done previously. [25] If the unstable region of the I-V curve extends past the NDR region, it is possible to underpredict the steepness of the conductance slope. However, we find that the average conductance slope across the unstable region is generally continuous with the slope before it, indicating that significant underprediction is likely not occurring.

3. The relative value of the conductance slope across different heterojunctions is far more useful for analysis than the absolute value of the slope for any given diode. Upon turning off the tunnel current, the diode behaves as a p-n junction, and so the tunnel current and conductance eventually becomes overwhelmed by the diode current. Consequently, the full swing in tunnel current cannot necessarily be seen. Also, when the diode current is significant, the fraction of the total measured current that is due to tunnelling decreases as the tunnel current drops, and so the calculated conductance slope will be less steep than it actually is. Therefore, the absolute value of the conductance slope is not overly telling of the sharpness one would obtain in a TFET, but the difference in conductance slopes between diodes is very telling as to which are giving a sharper transition.

## B. Tunnelling and Materials Defects

We expect electrically active defects and materials inhomogeneity to have an effect on the conductance slope, due to the presence of both 1) electronic states in the band gap that assist tunneling in the off state and 2) inhomogeneous band-alignment at the interface due to local band-bending in the region of defects as well as fluctuations in materials composition and strain. Figure 1a shows an example of an InAs/GaSb diode, with example defect states that provide a band-edge blurring via mechanism 1. In Figure 1b, an energy-band diagram is envisioned along a direction parallel to the junction in the region of an example defect, in this case, an acceptor-type deep level trap, similar to illustration by Wilshaw[39]. In addition to the trap state, there can be band-bending around it due to electric fields and/or local deformation of the lattice. The defect need not be a deep-level trap, rather, any defect concentration that is not uniformly spaced will result in nonuniform band-edges[40]. This means that such defects can alter the local band-alignment, as shown in Figure 1c, where the GaSb valence band right near the interface is shown and is assumed to not be affected by the defect for simplicity of illustration. Wherever the GaSb valence band lies below the InAs conduction band, tunneling can occur. As a voltage is applied to move the bands apart, tunneling will turn off at point C before point A. The band diagram here has been drawn for the case of no quantum confinement, although it is possible for the band-bending near the interface to act as shallow quantum wells for electrons and holes. Even if such quantum confinement occurs near the interface, the electric/fields and/or local lattice deformation will result in a change of both the 2-dimensional electron gas band-edges, as well as the energy of the confined state, and so the same result of nonuniform turn-off is to be expected. Figure 1d shows the result on conductance slope, where the slope is an average of the slope of every point. If such defects are concentrated enough, there will be significant averaging and hence, blurring of the turn-off. The effect of traps is also shown, via the presence of trap-assisted tunneling. Both of these effects will result in a less-steep conductance slope, while trap-assisted leakage will also result in a higher off-state current. This will be discussed more specifically in the results section.

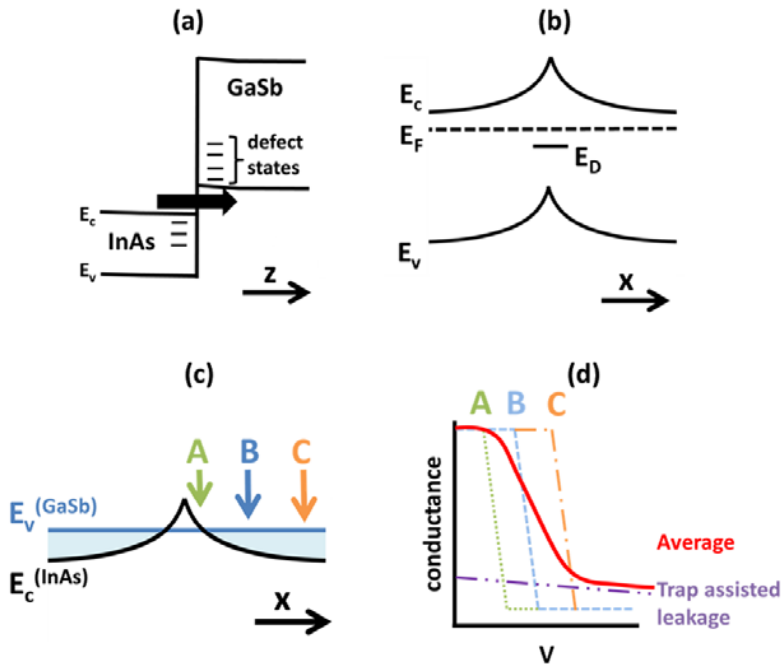

*Figure 1 a) Band diagram of InAs/GaSb diode. The arrow shows the direction of electron current due to tunnelling, and defect states are also shown near the interface, which can also act as sites for tunnelling. b) Band diagram of InAs along a direction parallel to the interface, showing a defect state and an example of band bending around the defect (the degree of band bending is not necessarily this large) c) Same diagram as b) with the valence band energy of GaSb near the interface shown. Light blue region represents where tunnelling can occur, and three reference points are marked as A, B, and C. d) Conductance-voltage curve showing sharpness of switching for current at each point A, B, and C, as well as a trap-assisted leakage current. The solid red curve shows the measured result, which is a weighted average of all of these conductance pathways, and the conductance slope is considerably less steep*

## C. Experimental Methods

InAs and GaSb thin films were grown in a custom-designed Thomas Swan/AIXTRON low pressure MOCVD reactor with a close coupled showerhead. Depending on the structure, growth was carried out on either epi-ready (100) p-GaSb substrates, or epi-ready (100) n-InAs substrates. TMGa and TMIn were used as the group-III precursors, and AsH$_3$ and TMSb were used as the group V precursors. Si$_2$H$_6$ was used as a Si doping source to obtain n-type doping ($1\times10^{17}$ cm$^{-3}$) for InAs and p-type doping ($1\times10^{17}$ cm$^{-3}$) for GaSb. All layers were doped to the same level across all devices, and electrical results were analyzed to ensure there were no changes in doping concentration between devices. The growth pressure was fixed at 100 torr and H$_2$ was used as the carrier gas. The growth temperature varied between 465ºC and 530 ºC. A homoepitaxial layer on the order of 100 nm in thickness was grown on the substrate before growing device layers. The thickness of InAs and GaSb layers varied between 60 and 120 nm. The default growth rate was 6 nm/min for GaSb and 12 nm/min for InAs. A 2s AsH$_3$ pulse always preceded the growth of a subsequent layer. For devices in which the intended top layer was InAs, a 20 nm n+ InAs contact layer was grown on top of the heterojunction, and for devices with GaSb as the top layer, a 20 nm p+ GaSb layer was grown, followed by a 60 nm InAs cap to prevent the p+ GaSb contact from being etched by the photoresist developer during diode fabrication.

Cross-section transmission electron micrscopy (TEM) was carried out on a JEOL 2011 microscope at an accelerating voltage of 200 kV to examine material quality. Relaxation was analyzed using glancing exit (224) reciprocal space maps on a Bruker D8 Advance high resolution x-ray diffractometer (HRXRD).

Circular mesa-diodes 12 µm in diameter were fabricated with top and bottom contacts using a self-aligned process where Ti (5 nm) / Pt (40 nm) / Au (120 nm) contacts were evaporated onto the top and bottom surface using a Tremscal FC2000 electron-beam evaporator. Top contacts were patterned by liftoff, and the resulting contact was used as an etch mask for wet-etching mesas, in which 2g:2mL:1mL citric acid:$H_2O$:$H_2O_2$ was used to etch InAs, and 1:5 $NH_4OH$:$H_2O$ was used to etch GaSb. This self-aligned method ensures the mesa is not larger than the contact, so that the external applied electric field is uniform across the junction. Current-voltage (I-V) curves were obtained on an Agilent B1500A Semiconductor parameter analyzer with Kelvin connections inside a Faraday cage. The conductance-voltage characteristic is examined with the series resistance removed by extracting the slope in the high-bias region of the I-V curve. Series resistance values ranged from $6.5\times10^{-6}$ to $1.2\times10^{-5}$ $\Omega\cdot cm^2$. Multiple measurements were taken on various devices, and all published I-V curves are representative of typical results and are consistently observed across different devices.

## III. Results and Discussion

### A. Intermixing, Dislocation Formation, and Effect of Growth Order

Figure 2a shows cross sectional TEM of InAs grown on GaSb at 530ºC, which is a commonly used temperature for InAs growth[27,28,31,33]. The interface appears to be very rough and numerous 60º threading dislocations appear to originate from the interface. The high concentration of misfit and threading dislocations can also be seen in plan view TEM as shown in Figure 2b. This defect density is not characteristic of small-lattice-mismatch materials such as InAs and GaSb (0.6%), indicating that these dislocations are not created by the small tensile strain in the growing InAs layer. Figure 2c shows a high-resolution TEM (HRTEM) image, where dark regions can be seen, indicating a difference in strain or composition. The inset shows a Burger's circuit indicative of a 90º (edge) dislocation in this region with a Burger's vector, $\vec{b} = -\frac{a}{2}[110]$, which is indicative of very large lattice mismatch (typically more than 4%) and could be a result of strain-induced island formation[41]. Furthermore, we have grown the InAs layer as thin as 10 nm, significantly below the critical thickness of InAs growth on GaSb[41], and find no improvement in defect density. All of these observations are indicative that this relaxation is not due to the small lattice mismatch between InAs and GaSb, but rather due to much larger strain that must be present at the interface. We believe that intermixing between the layers has led to intermediate compositions, such as Ga-rich $Ga_xIn_{1-x}As_ySb_{1-y}$, which can create enough strain to provide relaxation consistent with what is observed. The presence of intermediate composition has been seen in previous MBE and MOCVD growths[14,29,36]. Consistent with these previous reports, we expect the existence of a Ga carryover effect, where Ga incorporates in the above layer due to a proposed methyl-exchange reaction between TMIn and TMGa, and/or As-for-Sb swap, in which Sb exchanges with As and incorporates into the growing layer to lower surface energy, and also to favor the more stable Ga-As bond. In both of these processes, we expect a Ga-rich $Ga_xIn_{1-x}As_xSb_{1-x}$ region to form which is under high tensile stress and also puts compressive stress on the InAs layer growing above, and which is likely the source of dislocations.

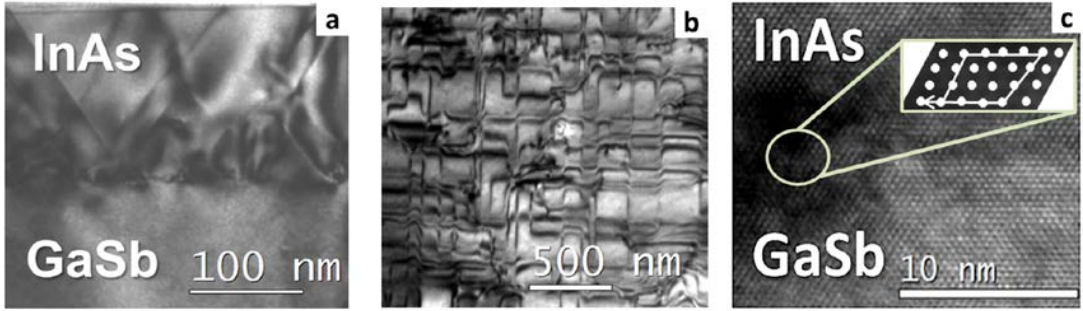

*Figure 2: a) Cross section TEM (220 diffraction condition) of InAs grown on GaSb at 530°C showing dislocations originating from the interface. b) Plan view TEM image of sample. c) Lattice-resolved TEM image of structure showing a dark region. Zoomed in portion shows a Burger's circuit drawn based on the identified lattice sites in the image.*

Both the As-for-Sb swap and Ga carryover processes are expected to be sensitive to the growth order of the interface, and hence should not occur when grown in the reverse order, with GaSb grown on top of InAs. This is because the As-for-Sb swap is driven heavily by the lower surface energy of Sb, and so it should only occur when Sb is in a lower layer in the structure and not the top layer. As for the Ga carryover effect, it is believed that the carryover occurs because Ga atoms in the lower layer react with the adsorbing TMIn to transfer methyl groups to the Ga atoms, which may allow it to become mobile and incorporate into the above film[42]. This methyl transfer is due to the stronger bond energy of the Ga-C bond than the In-C bond, and so the transfer should not occur between TMGa and In atoms in the scenario in which InAs is being grown on GaSb. Therefore, it is expected that these effects should not occur if GaSb is grown on top of InAs.

Figure 3 shows cross-section TEM for a structure grown in the reverse order, with GaSb as the top layer, for layer thicknesses of 80 nm and 380 nm. In both cases, there are no longer any visible threading dislocations. The 80 nm thick GaSb layer on InAs shows no visible interface defects under these TEM conditions. For the case of the 380 nm thick layer, it is typical of relaxation in a low-mismatch interface: misfit dislocations can be seen at the interface, indicating that there has been relaxation, but without the heavily defective interface and thread density seen for the InAs/GaSb structure. Figure 4 shows a comparison of strain relaxation for the InAs/GaSb structure and the two GaSb/InAs structures, via glancing exit (224) reciprocal space maps. For InAs/GaSb in Figure 4a, the reciprocal space position of the InAs (224) peak indicates >80% relaxation. For the reverse order in Figure 4b, it can be seen that at 80 nm thickness, the GaSb remains mainly strained with about 5% relaxation, while the 380 nm thick layer (Figure 4c) shows about 63% relaxation. The fact that the thread density remains significantly lower than the InAs/GaSb sample, even after significant relaxation, and that the GaSb can be grown considerably thick before relaxing and still not reach the same relaxation level as for InAs on GaSb, indicates that the intermixing-driven dislocation formation observed for InAs/GaSb heterostructures is not present when grown in the reverse order.

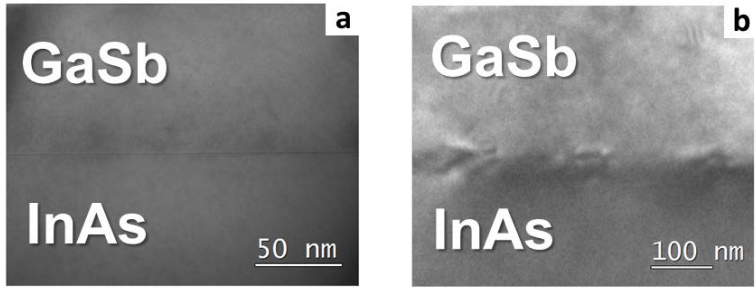

*Figure 3: Cross section TEM (220 condition) of a) 80 nm thick GaSb grown on InAs at 530°C, b) 380 nm thick GaSb on InAs (circular regions of striations in the bulk of the GaSb layer is milling-induced damage from the TEM sample preparation process)*

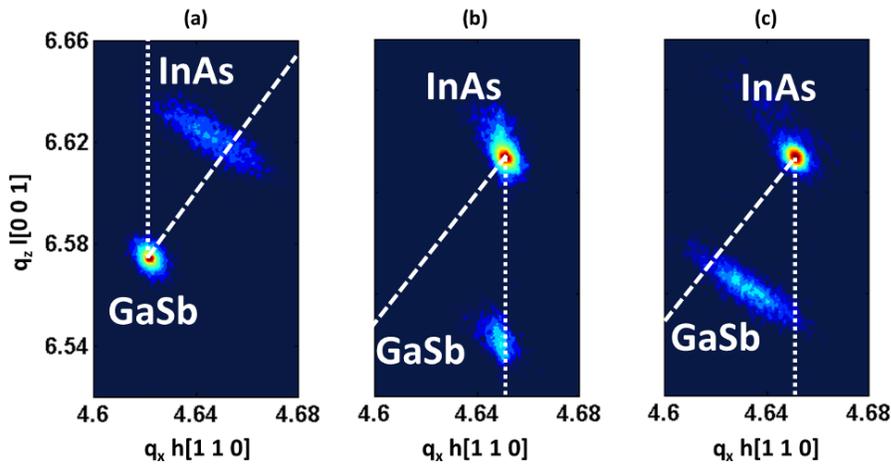

*Figure 4: (224) glancing exit reciprocal space maps of a) InAs/GaSb junction (82% relaxed) b) 80 nm GaSb / InAs (5% relaxed) c) 380 nm GaSb/InAs (63% relaxed). Short dashed line shows the line of strain, and long dashed line shows the line of relaxation*

The series-resistance-corrected I-V characteristics of the InAs/GaSb structure compared to the 80 nm thick GaSb/InAs structure is shown in Figure 5a, and the conductance-voltage characteristics are shown in Figure 5b. Negative differential resistance can be seen, and the two steep drops visible in the NDR region of I-V curves are due to instabilities in the test circuit resulting from the NDR, as is commonly observed in these types of measurements. It is worth reiterating that the conductance slope is taken as the average slope over the entire NDR region. The GaSb/InAs diode shows a much steeper conductance slope in the region of the bands uncrossing. This indicates that the heavily defective InAs/GaSb interface has caused a blurring of the turn-off, consistent with what was described in Figure 1. A less steep conductance slope can be seen, consistent with less uniformity, as well as a higher current beyond the steep portion, which appears consistent with more trap-assisted leakage, although the diode current becomes significant at higher forward biases, which can overshadow the leakage current. Further, while deep level traps would lead to a leakage current as depicted in Figure 1d, leakage through more shallow-level traps, if concentrated enough, would lower the slope throughout the entire region of band-uncrossing, not just at lower currents. Therefore, the main characteristic to extract from this data is the less-steep slope in the region in which the bands are uncrossing, which is attributable to a variety of energy states due to defects, and the local band-bending they cause.

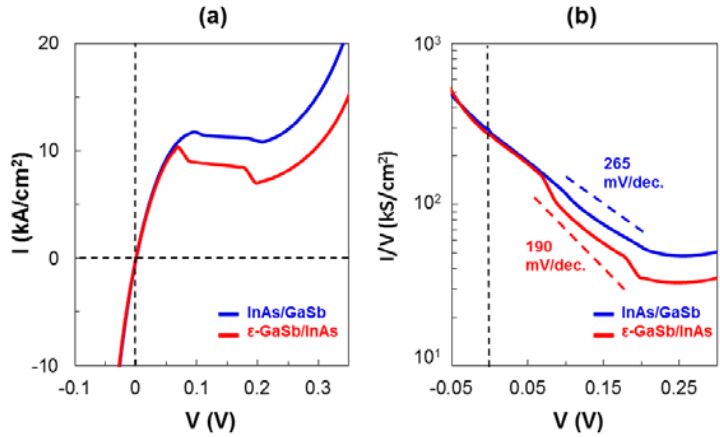

*Figure 5: a) Series resistance-corrected I-V curve and b) Absolute conductance-voltage curve derived from a), for InAs/GaSb and strained GaSb/InAs diodes*

This result does not indicate, however, which defects specifically cause this decrease in conductance slope steepness, as the InAs/GaSb interface has 60º and 90º misfit dislocations and many threading dislocations. Therefore, it is useful to compare the 80 nm GaSb/InAs diode to the case where the GaSb layer is grown to 380 nm thick and undergoes significant relaxation. In the latter case, the interface contains 60º misfit dislocations with a low threading dislocation density, as was seen in Figure 3b, but without many threads and edge dislocations as in the InAs/GaSb diode. Figure 6 shows the result: the conductance slope is significantly less steep and about the same as the InAs/GaSb diode, indicating that the loss of steepness is correlated to the presence of misfit dislocations. Theoretically, it can be proposed that the misfit dislocations are leading to energy states and local band bending in their region due either to 1) states associated specifically with the core of the dislocations, 2) states resulting from the local deformation of the lattice in the region of the dislocations or 3) states associated with point-defects, which tend to accumulate in the region of dislocations due to the difference in strain and charge. However, in reality, it is more likely to be a combination of 2) and 3), and mainly 3), as it has been shown previously that dislocations tend to getter many point defects which lead to states and lattice deformation which dominate the electrical characteristics[43], and that the dangling bonds in the core of most dislocations can reconstruct to passivate and eliminate the defect states that would be expected of a dislocation core[44]. This may explain why the InAs/GaSb diode, which in addition to having 60º dislocations also contained pure-edge 90º dislocations gave about the same conductance slope steepness: while a true 90º dislocation can have a different density of energy states than a 60º dislocation, both types attract point defects, and the effect of point defects overwhelms the much lower concentration of dangling bonds and reconstructed states.

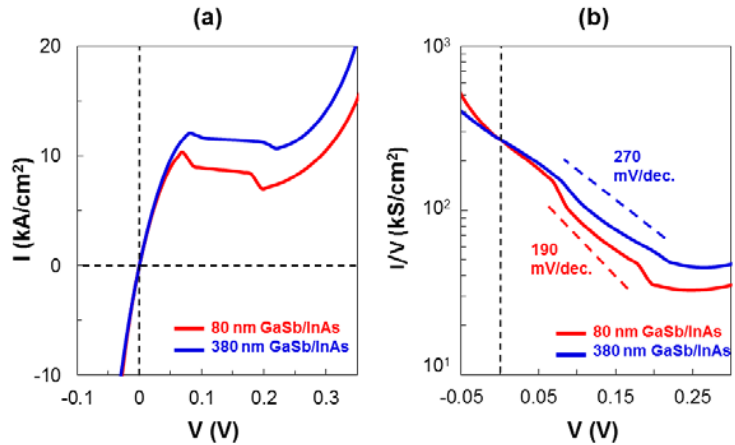

*Figure 6: a) Series resistance-corrected I-V curve and b) Absolute conductance-voltage curve for 80 nm thick and 380 nm thick GaSb on InAs diodes*

**B. Post-Growth Annealing, Point defects, and Interface Uniformity**

Given that the point defects gettered by dislocations and the resulting lattice-deformation are expected to lower the conductance slope steepness, it stands to reason that the steepness may be improvable via thermal annealing. This is because annealing can cause migration of point defects, which can allow for their annihilation. This migration can also act to average out the concentration to create a more uniform distribution. This would lead to fewer regions of largely different band alignment and hence, steeper switching in devices utilizing this interface.

We annealed both the high dislocation density InAs/GaSb and low dislocation density 80 nm thick GaSb/InAs at 600°C for 7.5 minutes. This was done in a rapid thermal annealer, with PECVD-deposited $SiO_2$ protective layers added to the top and bottom surface to prevent degassing, which was subsequently removed in a buffered oxide etch after annealing. Diodes were then fabricated on the annealed structures. We kept the anneal time short to avoid significant intermixing between the layers above what had already occurred during growth, and to avoid any relaxation in the still strained GaSb/InAs diode.

Figure 7a and b show the I-V curve and conductance-voltage curve of the InAs/GaSb device with and without the annealing step. It can be seen that there is a noticeable increase in the steepness of the conductance slope after annealing, which is indicative that in both cases, point defects are being eliminated and/or more evenly distributed.  It is clear, however, that the annealed device still has many electrically active defects, as the steepness after annealing is still not as good as the device without dislocations. We do not expect that there was much intermixing between the layers during the annealing, as this would lead to a noticeable decrease in current in reverse bias due to a lower tunnel probability, and would likely lead to a lower conductance slope as well. But it can be seen that the current and conductance in reverse bias, where tunneling is occurring, has not changed after annealing.

Even without a substantial concentration of misfit dislocations, there is a significant improvement in conductance slope, as is shown in Figure 7c and 7d, for the annealing of the GaSb/InAs structure which is almost completely strained.  This is likely an indication that all samples contain point defects. The presence of the misfit dislocations at the interface tends to getter these point defects and increase their concentration. Hence, the conductance slope is made steeper by annealing dislocated samples, enhanced further by preventing dislocations, and enhanced even further by annealing undislocated samples. Each enhancement is due to fewer point defect states and more interface uniformity.  Again in this case, the electrical results do not indicate increased intermixing. Furthermore, we confirmed via HRXRD that there was no additional relaxation of the GaSb layer after annealing, consistent with the much steeper slope.

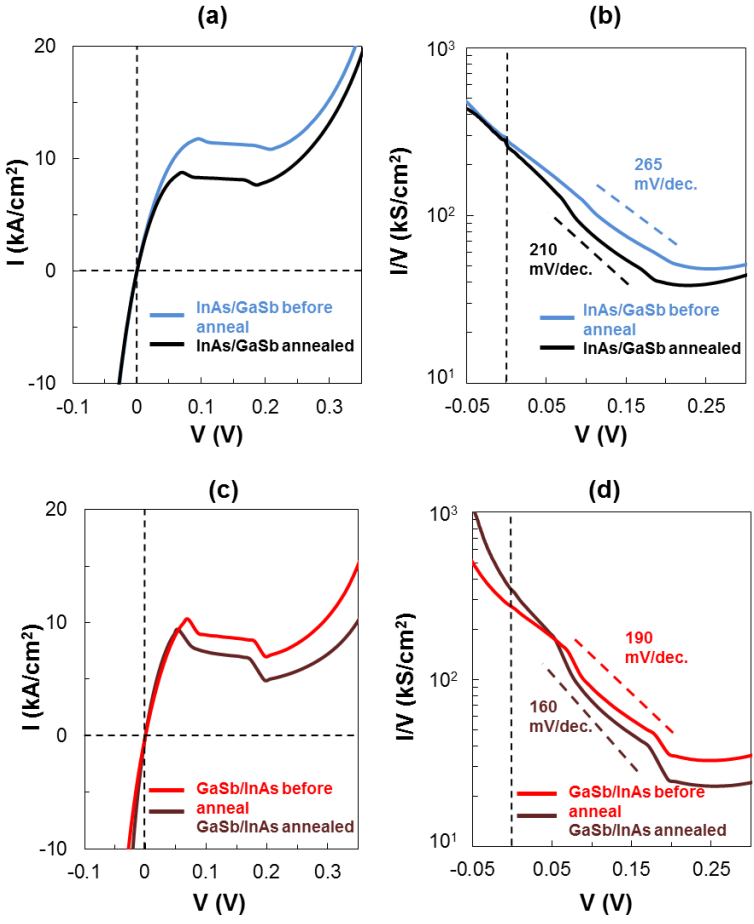

*Figure 7: a) Series resistance-corrected I-V curve and b) Absolute conductance-voltage curve InAs/GaSb diodes before and after annealing at 600 ⁰C. c) and d): Equivalent comparison for 80 nm thick GaSb/InAs diodes*

**C. Use of Strain to Suppress Defect Formation**

Thus far, it has been shown that GaSb can easily be grown strained on InAs, and this prevents misfit dislocation formation and the less-steep conductance slope that results from it. However, at the normal growth temperature of 530 °C, it is not possible to grow the InAs strained, as it relaxes immediately due to the much larger strain of intermediate compositions. However, since the GaSb can be so easily strained, it is possible to "pseudo lattice match" the InAs to the GaSb to prevent relaxation, by keeping the GaSb strained on the InAs lattice constant, and then growing the InAs layer on top of the GaSb layer without lattice mismatch. To test if this is effective, we grew InAs layers on both the strained and relaxed GaSb layers to determine if keeping the GaSb underlayer strained provides any improvement. Figure 8a shows a cross-section TEM image for InAs grown on the 380 nm relaxed GaSb layer, and is heavily dislocated in the same manner as when InAs is grown on GaSb substrates. However, as shown in Figure 8b, when InAs is grown on the strained GaSb layer, it has no observable defects in cross-section TEM. This seems to indicate that growing the InAs on GaSb that has been strained to the InAs lattice constant prevents either the intermixing that leads to high strains, or the relaxation of these high strains. It is possible that that intermixing is reduced because lattice-matching could prevent island formation and so the entire surface is quickly covered in InAs, making As-for-Sb exchange and Ga carryover more difficult. Alternatively, it is possible that the intermixing still occurs, but the film mismatch is required for relaxation. This is conceivable since the intermixing by exchange or carryover would have a shrinking driving force as strain begins to buildup, and it is unlikely that it would continue beyond a point where enough strain energy was built up for relaxation mechanisms to activate. Hence, we would expect the diffusion to lead to unrelaxed strain. Additional strain would then be required to lead to relaxation and island formation. Hence, the film mismatch is needed to bring the strain energy to the required minimum for strain relaxation to be energetically favorable, and removing the lattice mismatch between the GaSb and InAs prevents the strain relaxation from ever becoming energetically favorable.

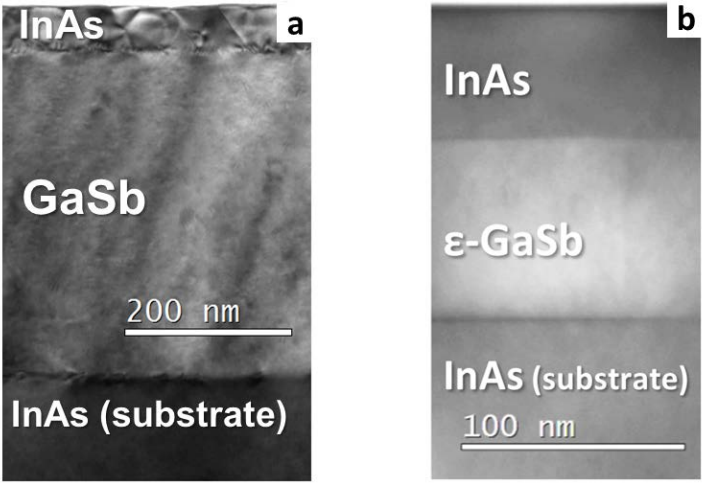

*Figure 8: TEM images of: a) InAs grown on 380 nm thick (63% relaxed) GaSb, b) InAs grown on 80 nm thick strained GaSb (images are at different magnifications to show all layers. Top InAs layer is the same thickness for both)*

Diodes were fabricated on both of these samples. However, since there are now two interfaces (as can be seen in Figure 8) - the main InAs on GaSb tunnel interface, as well as the lower interface for the GaSb layer on the InAs substrate - great care was taken to prevent the lower interface from affecting the I-V curve. This was done by regrowing n+-InAs on the InAs substrate, and growing the first half of the GaSb film as Si p+ doped. This makes the lower tunnel interface much more conductive. Furthermore, the GaSb was not etched all the way through the layer, to allow for room for current spreading in the GaSb layer, making the effective area of the lower junction larger than that of the main tunnel junction. Figure 9a shows the I-V curve for both cases, and Figure 9b shows the conductance-voltage curve. The results are consistent with the rest of the study, in that the dislocated InAs on relaxed-GaSb shows the same less steep conductance slope (refer back to Figure 5b for a comparison). The InAs/strained-GaSb device, on the other hand, shows a sharper conductance slope, due to its lack of dislocations. It is worth noting that for these plots, while the series resistance was removed in the standard manner of using the slope at high voltage, the variable resistance of the lower interface tunnel interface cannot be removed this way, since it is small at high voltage and becomes larger at small forward bias and increasing reverse bias. This is likely the origin of the loss of steepness through the origin for both diodes in Figure 9b. Since this variable resistance was the same for both diodes, the comparison is still fair.

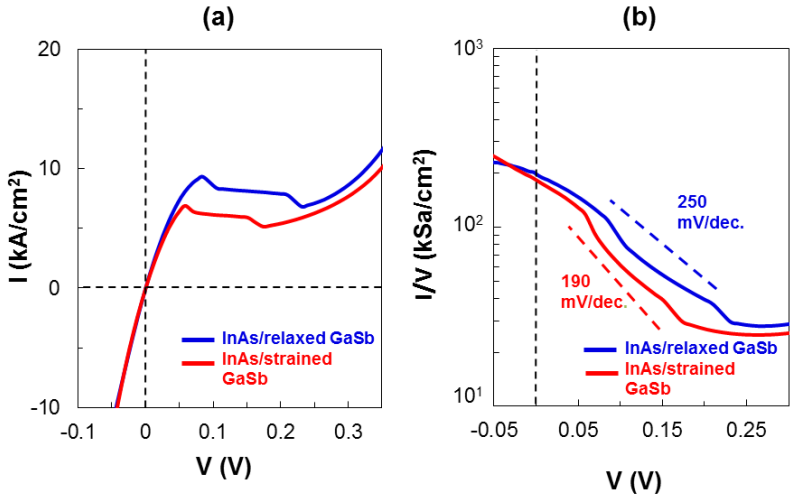

*Figure 9: a) Series resistance-corrected I-V curve and b) Absolute conductance-voltage curve for InAs grown on 80 nm thick (strained) and 380 nm thick (63% relaxed) GaSb*

**D. Use of Growth Temperature to Suppress Defect Formation**

While growing in reverse-order or with a strained GaSb underlayer prevents dislocation formation, these structures could lead to difficulty in later fabricating a three-terminal device. Another possibility, given that dislocations ultimately originate from intermixing, would be to lower the growth temperature of the InAs layer to prevent this intermixing. We attempted this by reducing the growth temperature from 530°C after growing the GaSb layer, down to 465 °C for InAs growth. Cross section TEM results are shown in Figure 10a for a structure with InAs grown at 465°C, which can be compared to the original 530 °C growth in Figure 2. Figure 10b shows a plan view TEM of the heterojunction with the InAs grown at 465 ºC. There is an improvement in interface quality and a drop in threading dislocation density as temperature decreases, indicating less intermixing and dislocation formation. Figure 11a and b show a glancing exit (224) reciprocal space map for the InAs layer grown at 530 °C and 465°C. The layer grown at 465 °C is <1% relaxed.

The results of this study indicate that higher quality interfaces can be obtained by lowering the temperature of the InAs layer growth to prevent intermixing by either As-for-Sb exchange, or Ga carryover, or both. This may be due to the fact that the lower temperature provides less kinetic energy to surmount the activation energy barriers associated with swap and carryover. Lowering the temperature can also allow the InAs layer to coalesce earlier in the growth, leading to InAs/InAs growth sooner and a 2D growth mode sooner. It has been reported previously that growing InAs on GaSb in MBE above 460ºC resulted in island formation due to the strain of an intermediate Ga-rich layer(s)[37]. It is possible that at higher temperatures, the Ga-rich intermediate material causes enough strain that the InAs layer forms islands, which then leaves the GaSb layer underneath exposed in some areas for a longer period of time, allowing further exchange to occur between the Sb in the GaSb layer, and the incoming As flux. By preventing island formation, there would be no exposed GaSb regions after growth initiated, and the exchange process would become self-limiting.

To verify that the lower dislocation density at lower temperature was a result of decreased intermixing, and not simply due to increased barriers to dislocation formation, we grew a sample where a 40 nm thick InAs layer was grown at 465 ºC, followed by an additional 50 nm grown while ramping the temperature from 465 ºC to 530 ºC, and then 50 nm of growth at 530 ºC. Figure 11c shows the (224) reciprocal space map, showing a heavily strained InAs layer with little relaxation. This indicates that intermixing has been suppressed. Had the lower temperature only suppressed dislocation formation, it is likely that ramping back up to 530 ºC would have led to dislocation formation and relaxation. Instead, we observed a similar structure to the 465 ºC structure, implying that the lower temperature likely suppresses the exchange and/or carryover. Then, when the temperature is ramped back up, this cannot enhance the exchange and carryover processes, since the interface is already buried, and both of these processes require a free surface to mediate the intermixing.

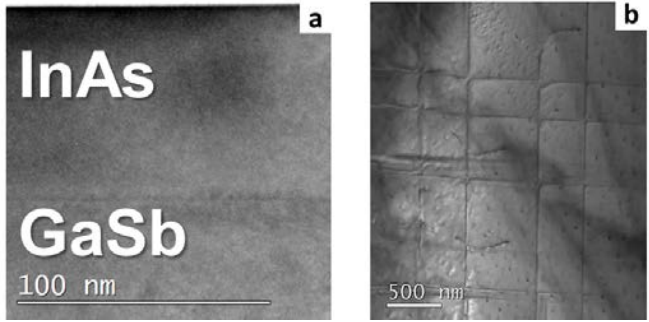

*Figure 10: a) Cross section TEM image of InAs grown at 465°C on GaSb, b) Plan view TEM of the sample (small round defects are milling induced damage)*

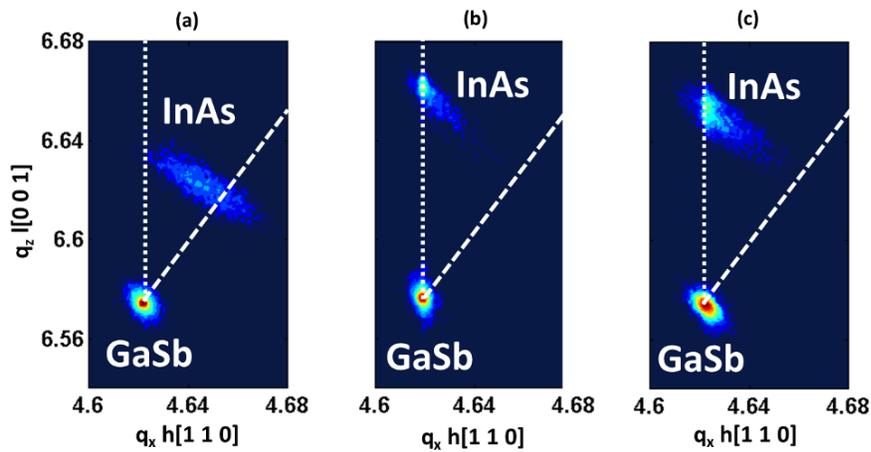

*Figure 11: (224) glancing exit reciprocal space maps of a) InAs/GaSb junction grown at 530 $^{0}$C (82% relaxed) b) InAs grown at 465 $^{0}$C on GaSb (<1% relaxed) c) InAs grown at 465 $^{0}$C and*

*then ramped up to 530 °C while growing. Short dashed line shows the line of strain, and long dashed line shows the line of relaxation*

Figure 12a and 12b compare the electrical results of three devices: 1) a device with the InAs layer grown entirely at 465 °C, 2) a device with 40 nm InAs grown at 465 °C and then ramped up to 530 °C while growing, and 3) a device with the InAs layer grown at 465 °C and then the entire epitaxial structure annealed at 600 °C for 7.5 minutes, using the anneal procedure described previously. Device (1) shows a change in conductance, but with extremely small steepness which levels off at a fairly flat slope afterwards, indicating that significant tunneling is still occurring either through defect states or regions where the band overlap is still large. For device (2), growing with a ramp in growth temperature to 530 °C provides some improvement, but minor. However, for device (3), when the structure is annealed at 600 °C, the conductance slope steepness improves greatly, giving a conductance slope about the same as the annealed dislocation-free GaSb/InAs device, consistent with the fact that this device indeed is also a dislocation-free annealed device. The phenomenon here appears to be the same as that for annealing the GaSb/InAs structure, in that the annihilation of point defects and/or averaging of fluctuations in band alignment have increased the steepness of the conductance slope. However, it appears in this case that the steepness was much weaker before annealing. This makes sense since we expect a growth temperature below the optimized growth temperature of 530 °C to introduce more point defects into the layer, due to a variety of possible mechanisms such as lack of surface diffusion during growth resulting in defect concentrations above the equilibrium value, greater carbon-incorporation due to incomplete precursor decomposition, increased impurity incorporation due to lower growth rate, and a shift in the equilibrium stoichiometry of the compound via a change in partial pressure of the In and As, via a change in the precursor decomposition rate. Also, since after the GaSb layer is grown, the temperature must be ramped down to 465 °C before the InAs can be grown, there is the opportunity to incorporate impurities at the surface. Despite all of these issues, however, dropping the InAs growth temperature to such a low value is still beneficial, since the issues it introduces can be resolved by annealing, and the net result is a major improvement over growing at 530 ºC. Again, we expect minimal intermixing due to annealing given the lack of change in conductance in the tunnel region (reverse bias), and confirmed with HRXRD that annealing led to a negligible amount of relaxation.

Figure 12c compares this annealed device to the device grown at 530 ºC annealed and unannealed. This agrees with and reiterates the concept stated in Section B: there is improvement in steepness of conductance slope in annealing dislocated samples, further improvement in a sample with no dislocations, and even further improvement in an annealed sample with no dislocations.

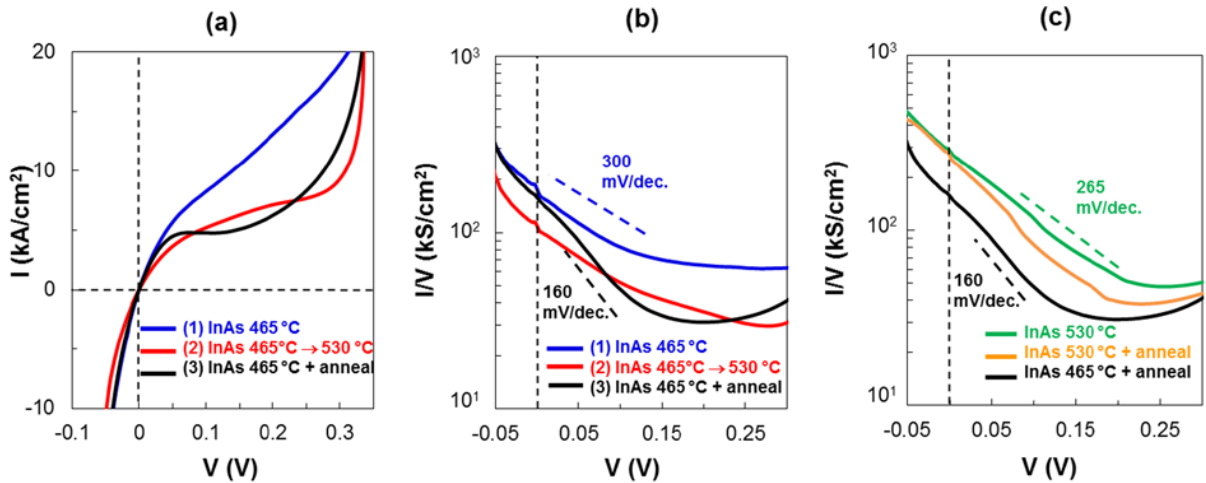

*Figure 12: a) I-V curve and b) Absolute conductance-voltage curve for (1) InAs grown at 465 ⁰C, (2) 465 ⁰C with a ramp to 530 ⁰C, and (3) 465 ⁰C and then the entire structure annealed at 600 ⁰C. c) Comparing the annealed sample to growth at 530 ⁰C with and without annealing*

## IV. Conclusion

We have identified point and line defects as having a major effect on the steepness of the conductance slope of InAs/GaSb tunnel diodes. For line defects, misfit dislocations dominate and lead to a major decrease in conductance slope steepness. This is presumed to be due to gettering of point defects leading to energy states and bend bending that allow for trap-assisted leakage and nonuniform band alignment across the interface. This is unavoidable for InAs/GaSb structures grown at 530°C, since the material relaxes instantly due to the buildup of strain from intermediate compositions from intermixing. This can be circumvented by growing in the reverse order, to prevent the intermixing, or growing on strained-GaSb, so that the InAs layer is lattice-matched. It can also be avoided by lowering the growth temperature of the InAs layer to prevent intermixing.

Even without dislocations, point defects still affect the performance of the devices, and all devices respond positively to annealing, as it lowers the concentration of these defects.

Future work will focus on the addition of barrier layers at the interface, including AlSb, in order to assess if there is any effect by blocking leakage mechanisms. We further plan to assess alloyed systems in which the band alignment can be brought closer to the boundary of type-II/type-III band alignment.

## V. Acknowledgements

This work was supported by the Center for Energy Efficient Electronics Science (E3S) (NSF Award 0939514). The author also thanks the Natural Sciences and Engineering Research Council of Canada for a Postgraduate M Scholarship which was used while carrying out this research. We also thank Nathan Martin, an undergraduate intern funded by E3S who assisted with TEM and XRD measurements, as well as Kunal Mukherjee, Tim Milakovich, Adam Jandl,

Prithu Sharma, Roger Jia, Chris Heidelberger, Rushabh Shah, Mayank Bulsara, Sapan Agarwal, Jared Carter, Jamie Teherani, Winston Chern, Tao Yu, Andrew Carlin, Andrew Malonis, and Michael Burek for helpful discussions.## VI. References

[1] A. C. Seabaugh and Q. Zhang, Proceedings of the IEEE **98,** 2095 (2010).

[2] A. M. Ionescu and H. Riel, Nature **479,** 329 (2011).

[3] O. M. Nayfeh, C. N. Chleirigh, J. Hennessy, L. Gomez, J. L. Hoyt, and D. A. Antoniadis, Electron Device Letters, IEEE **29,** 1074 (2008).

[4] S. O. Koswatta, S. J. Koester, and W. Haensch, in *Electron Devices Meeting (IEDM)*, 2009 (IEEE), p. 1.

[5] J. Knoch and J. Appenzeller, Electron Device Letters, IEEE **31,** 305 (2010).

[6] S. O. Koswatta, S. J. Koester, and W. Haensch, IEEE Transactions on Electron Devices **57,** 3222 (2010).

[7] L. Wang, E. Yu, Y. Taur, and P. Asbeck, Electron Device Letters, IEEE **31,** 431 (2010).

[8] Y. Lu, G. Zhou, R. Li, Q. Liu, Q. Zhang, T. Vasen, S. D. Chae, T. Kosel, M. Wistey, H. Xing, A. Seabaugh, and P. Fay, Electron Device Letters, IEEE **33,** 655 (2012).

[9] B. M. Borg, K. A. Dick, B. Ganjipour, M. E. Pistol, L. E. Wernersson, and C. Thelander, Nano Letters **10**, 4080 (2010).

[10] R. Li, Y. Lu, G. Zhou, Q. Liu, S. D. Chae, T. Vasen, W. S. Hwang, Q. Zhang, P. Fay, T. Kosel, M. Wistey, H. Xing, and A. Seabaugh, Electron Device Letters, IEEE **33,** 363 (2012).

[11] R. Li, Y. Lu, S. D. Chae, G. Zhou, Q. Liu, C. Chen, M. Shahriar Rahman, T. Vasen, Q. Zhang, P. Fay, T. Kosel, M. Wistey, H. Xing, S. Koswatta, and A. Seabaugh, Physica Status Solidi (c) **9,** 389 (2012).

[12] G. Zhou, R. Li, T. Vasen, M. Qi, S. Chae, Y. Lu, Q. Zhang, H. Zhu, J. Kuo, T. Kosel, M. Wistey, P. Fay, A. Seabaugh, and H. Xing, in *Electron Devices Meeting (IEDM)*, 2012, p. 32.6.128

13   A. W. Dey, B. M. Borg, B. Ganjipour, M. Ek, K. A. Dick, E. Lind, P. Nilsson, C. Thelander, and L.-E. Wernersson, in *Device Research Conference (DRC), 2012 70th Annual*, 2012 (IEEE), p. 205.

14   Y. Zhu, N. Jain, S. Vijayaraghavan, D. Mohata, S. Datta, D. Lubyshev, J. Fastenau, W. Liu, N. Monsegue, and M. Hudait, Journal of Applied Physics **112,** 024306 (2012).

15   D. Collins, E. Yu, Y. Rajakarunanayake, J. Soderstrom, D. Z. Y. Ting, D. Chow, and T. McGill, Applied Physics Letters **57,** 683 (1990).

16   J. Chen, L. Yang, M. Wu, S. Chu, and A. Cho, Journal of Applied Physics **68,** 3451 (1990).

17   J. Chen, M. Wu, L. Yang, and A. Cho, Journal of Applied Physics **68,** 3040 (1990).

18   R. Beresford, L. Luo, K. Longenbach, and W. Wang, Applied Physics Letters **56,** 952 (1990).

19   J. Schulman and D. Chow, Electron Device Letters, IEEE **21,** 353 (2000).

20   C. Burrus, IEEE Transactions on Microwave Theory and Techniques **11,** 357 (1963).

21   Z. Zhang, R. Rajavel, P. Deelman, and P. Fay, Microwave and Wireless Components Letters, IEEE**, 21**, 267 (2011).

22   D. Pawlik, B. Romanczyk, P. Thomas, S. Rommel, M. Edirisooriya, R. Contreras-Guerrero, R. Droopad, W. Loh, M. Wong, and K. Majumdar, in *Electron Devices Meeting (IEDM), 2012 IEEE International*, 2012 (IEEE), p. 27.1. 1.

23   B. Romanczyk, P. Thomas, D. Pawlik, S. Rommel, W. Loh, M. Wong, K. Majumdar, W. Wang, and P. Kirsch, Applied Physics Letters **102,** 213504 (2013).

24   L. Luo, R. Beresford, and W. Wang, Applied Physics Letters **55,** 2023 (1989).

25   S. Agarwal, E. Yablonovitch, IEEE Transactions on Electron Devices, **61**, 1488 (2014)

26   K. Taira, I. Hase, and H. Kawai, Electronics Letters **25,** 1708 (1989).

27   Y. Huang, J. H. Ryou, R. Dupuis, V. D'Costa, E. Steenbergen, J. Fan, Y. H. Zhang, A. Petschke, M. Mandl, and S. L. Chuang, Journal of Crystal Growth **314**, 92 (2011):

28   X. Zhang, J. Ryou, R. Dupuis, C. Xu, S. Mou, A. Petschke, K. Hsieh, and S. Chuang, Applied Physics Letters **90,** 131110 (2007).

29 G. Booker, P. Klipstein, M. Lakrimi, S. Lyapin, N. Mason, R. Nicholas, T. Y. Seong, D. Symons,T. Vaughan, and P. Walker, Journal of Crystal Growth **145,** 778 (1994).

30 Y. Su, C. Lin, S. Chen, J. Chang, and D. Jaw, Journal of Applied Physics **81,** 7529 (1997).

31 X. Zhang, J. Ryou, R. Dupuis, A. Petschke, S. Mou, S. Chuang, C. Xu, and K. Hsieh, Applied Physics Letters **88,** 072104 (2006).

32 M. Lakrimi, R. Martin, N. Mason, R. Nicholas, and P. Walker, Journal of Crystal Growth **110,** 677 (1991).

33 Y. Huang, J. H. Ryou, R. D. Dupuis, D. Zuo, B. Kesler, S. L. Chuang, H. Hu, K. H. Kim, Y. T. Lu, K. C. Hsieh, and J. M. Zuo, Applied Physics Letters **99,** 011109 (2011).

34 D. Lackner, O. Pitts, S. Najmi, P. Sandhu, K. Kavanagh, A. Yang, M. Steger, M. Thewalt, Y. Wang, D. McComb, C. Bolognesi, and S. Watkins, Journal of Crystal Growth **311,** 3563 (2009).

35 Y. Zhu, N. Jain, S. Vijayaraghavan, D. Mohata, S. Datta, D. Lubyshev, J. Fastenau, A. K. Liu, N. Monsegue, and M. Hudait, Journal of Applied Physics **112,** 094312 (2012).

36 Y. Zhu, D. Mohata, S. Datta, and M. Hudait, IEEE Transactions on Device and Materials Reliability **14**, 245 (2014).

37 Q. Xie, J. Van Nostrand, J. Brown, and C. Stutz, Journal of Applied Physics **86,** 329 (1999).

38 R. Kaspi, Journal of Crystal Growth **201,** 864 (1999).

39 P. Wilshaw and G. Booker, Structure and Properties of Dislocations in Semiconductors," Isz. Akad Nauk, USSR (1987).

40 D. Redfield, Physical Review **130,** 914 (1963).

41 E. Fitzgerald, Materials Science Reports **7,** 87 (1991).

42 V. Donnelly and J. McCaulley, Surface Science **235,** L333 (1990).

43 C. Kisielowski, J. Palm, B. Bollig, and H. Alexander, Physical Review B **44,** 1588 (1991).

44 J. Hornstra, Journal of Physics and Chemistry of Solids **5,** 129 (1958).